# A spatially explicit impact assessment of road characteristics, road-induced fragmentation and noise on bird species in Cyprus

Konstantinos Konstantopoulos[1], Aristides Moustakas[2], Ioannis N. Vogiatzakis[1]*
[1]*School of Pure & Applied Sciences, Open University of Cyprus, PO Box 12794, 2252, Nicosia, Cyprus*
[2] *Natural History Museum, University of Crete, Heraklion, Greece*
*Corresponding author: ioannis.vogiatzakis@ouc.ac.cy

**Abstract**
The rapid increase of transportation infrastructure during the recent decades has caused a number of effects on bird species, including collision mortality, habitat loss, fragmentation and noise. This paper investigates the effects of traffic noise and road-induced fragmentation on breeding bird richness in Cyprus. Cyprus, situated along one of the main migratory routes for birds, has a rich and diverse avifauna threatened by an ever-expanding road network and a road density among the highest in Europe. In this first island-wide study we used data from 102 breeding birds recorded in 10 km × 10 km grid cells. Within every cell we calculated road traffic noise and eight road-related properties. Most of the grid cells are subject to intense fragmentation and traffic noise with combined impact 'hotspots' located even within protected areas (such as Cape Greco, and the Troodos Massif). Results from variance partitioning indicated that road-related properties (total road extent and road length) accounted for a combined 59% of variation in species richness, followed by fragmentation-related properties and noise properties. The study posits the need for further in-depth research on the effects of road networks on birds, and road construction, particularly in protected areas within Mediterranean islands.

**Introduction**

With the global human population being higher than ever before and increasing exponentially, together with unprecedented rates of mobility of humans and goods, there are more roads than ever before. Roads fragment natural habitats, introduce additional noise and dust in the environment (Taiwo et al. 2017), and diseases via human mobility (Bertuzzo et al. 2016), among other impacts on the environment such as hazardous materials transportation (Wang et al. 2016). The environmental impact of roads is particularly acute in island ecosystems as their size is restricted, but often the industrial development within them is not. Bird populations around the world are declining (Birdlife International 2008a) at a fast rate. In North America, over 20 common species have declined more than 50% within the last 40 years (Butcher and Niven 2007; Birdlife International 2008b). Furthermore, more than half of migratory neotropical species are subject to decline, with underlying reasons not yet fully understood (Butcher and Niven 2007; Birdlife International 2008c). In Western Europe, in addition to climate change, farmland bird population declines have also been attributed to agricultural intensification. (Donald et al. 2001) and land- use change (Butler et al. 2010), although this is contested for forest birds in some geographical areas (see review by Reif 2013). Part of this decline has been attributed to direct and indirect road-induced effects, mainly due to fragmentation and traffic noise (Benitez-Lopez et al. 2010; Kociolek et al. 2011; Ortega 2012; Torres et al. 2016).

Habitat fragmentation, defined as the process by which a large area of habitat is converted into smaller patches, isolated from each other and from the landscape matrix (Fahrig 2003; 2017), is – together with habitat loss – a significant threat to worldwide biodiversity (Fischer and Lindenmayer 2007). In European landscapes, fragmentation due to increased urban sprawl and continuous expansion of transportation infrastructure in the countryside have amplified existing problems (Torres et al. 2016), yet the ecological impacts are not fully understood (Kociolek et al. 2011). This process is more obvious in urbanised or highly exploited areas, where fragmentation is caused by the connection of urban areas with transportation infrastructure, e.g. roads and railways (Van Der Ree, Smith and Grilo 2015). Many of the negative effects of roads on other vertebrates (e.g. mortality, fragmentation of habitats, audio-visual disturbance, chemical pollution) also impact birds populations (Forman et al. 2003; Jacobson 2005), although benefits for birds have also been reported (Forman 2000; Huijser and Clevenger 2006; Reijnen and Foppen 2006; Lambertucci et al . 2009).

One of the main ecological impacts of road-induced fragmentation is on habitat connectivity, endangering long-term populations' persistence. Evidence suggests that birds are influenced negatively by road density, because of direct and indirect impacts such as road mortality, noise and habitat loss (Trombulak and Frissell 2000; Gunderson et al. 2005; Fahrig and Rytwinsky 2009). However, additional factors such as road maintenance and repairs may deteriorate habitat quality and destroy nesting sites resulting in reduced population viability (Catlin and Rosenberg 2006).

Roads provide human access to wildlife habitats and facilitate the spread of invasive species (Arianoutsou et al. 2010), while subdivision and isolation of subpopulations may interrupt metapopulation dynamics (Forman et al., 2003) and reduce genetic diversity (Forman and Alexander 1998; IUCN 2001). Landscape fragmentation also increases the risk of population extinction, because isolated populations tend to be more sensitive to natural stress factors, e.g. natural disturbances, adverse weather conditions, wildfires and diseases, reducing their adaptability. Landscape fragmentation is the main reason for the rapid decline of many wild populations. Since landscape fragmentation can lead to the destruction of established ecological connections between adjacent areas of landscape (Haber 1993; Jaeger et al. 2005), whole communities and ecosystems can be affected. The possibility that two individuals of the same species can find each other is a prerequisite for the conservation of animal populations (e.g. due to the need for genetic exchange between populations and for recolonization of vacant habitats). There may well be additional consequences for which our knowledge is still limited, including the response time of wild populations to these such impacts and the effects upon ecological communities (e.g. overlapping effects).

The extension of paved roads and the consequent increase in traffic speed and volume in these roads are contributing factors to this decline (Ritters and Wickham 2003). Traffic noise has the most widespread and significant indirect effect on birds (Reijnen et al. 1995). Noise probably causes reduction in population' frequencies (i.e. acoustic masking) reported for various bird species near roads (Reijnen and Foppen 2006; Patricelli and Blickley 2006). In grasslands, the acoustic masking of noise attenuates less rapidly than in forests (Forman et al. 2002), presumably due to vegetation characteristics. In addition to traffic volume and noise impacts, there may be correlated (cumulative) impacts of noise, habitat loss, fragmentation (Forman and Deblinger 2000) and edge effects (Habib et al. 2007).

Cyprus is an island with special importance for ornithology because it is situated along one of the main migratory routes for birds, which has resulted in a rich and diverse avifauna. At the same time, it is among the countries in Europe with the highest road densities (Zomeni and Vogiatzakis 2014) and where the road network is expanding (Statistical Service of Cyprus data). This is the first large scale/island-wide study looking at the combined effects of traffic noise and road induced fragmentation on bird richness. Therefore, the objectives were to: a) map the extent of road traffic noise and road induced fragmentation on the island and; b) quantify the effects of road-related properties, road induced-fragmentation, and noise related properties on breeding birds' richness.

**Methods**

*Study Area*

Cyprus is the third largest island in the Mediterranean (9.251 km$^2$) with a climate which is characterized by extended warm and dry summers and mild, rainy winters. The island has high endemism in various taxonomic groups including birds. Situated at the southeastern edge of the Mediterranean, along one of the main migratory routes between Europe and Africa, it is important for migratory species breeding and overwintering. Cyprus is considered one of the most important areas for avifauna globally (Charalambidou et al. 2016). There are 397 recorded bird taxa until now including two endemic species (*Sylvia melanothorax, Oenanthe cypriaca*) and four endemic subspecies (*Otus scops cyprius, Parus ater Cypriotes, Certhia brachydactyla dorotheae, Garrulus glandarius glaszneri*). Among the breeding species, there are 53 permanent residents, of which 42 nest on the island regularly and 18 that have nested at least once.

*Datasets*

Bird data were provided by Birdlife Cyprus. These included grid data for the distribution of 102 breeding birds of Cyprus (resolution 10 km x 10 km), which were based on the 2013 field survey. These data have been collected by observers and volunteers following the European Atlas of Breeding Birds Methodology ([http://www.ebcc.info/atlas.html)](http://www.ebcc.info/atlas.html)., with total species presence aggregated per grid cell. Out of 137 grid squares covering the island only 75 were retained for the analysis herein (Fig 1). Those grid squares which were under-surveyed, mainly in the north, as well as those within major urban centres or with their centroid on the sea were excluded. Road and traffic data for the Annual Census of Traffic of 2012 was provided by the Department of Public Works of the Ministry of Transport, Communication and Works of Republic of Cyprus ([www.mcw.gov.cy/pwd](www.mcw.gov.cy/pwd)). In particular, the dataset contained information on road type and length, direction, traffic volume per type of vehicle and average traffic volume per road segment.

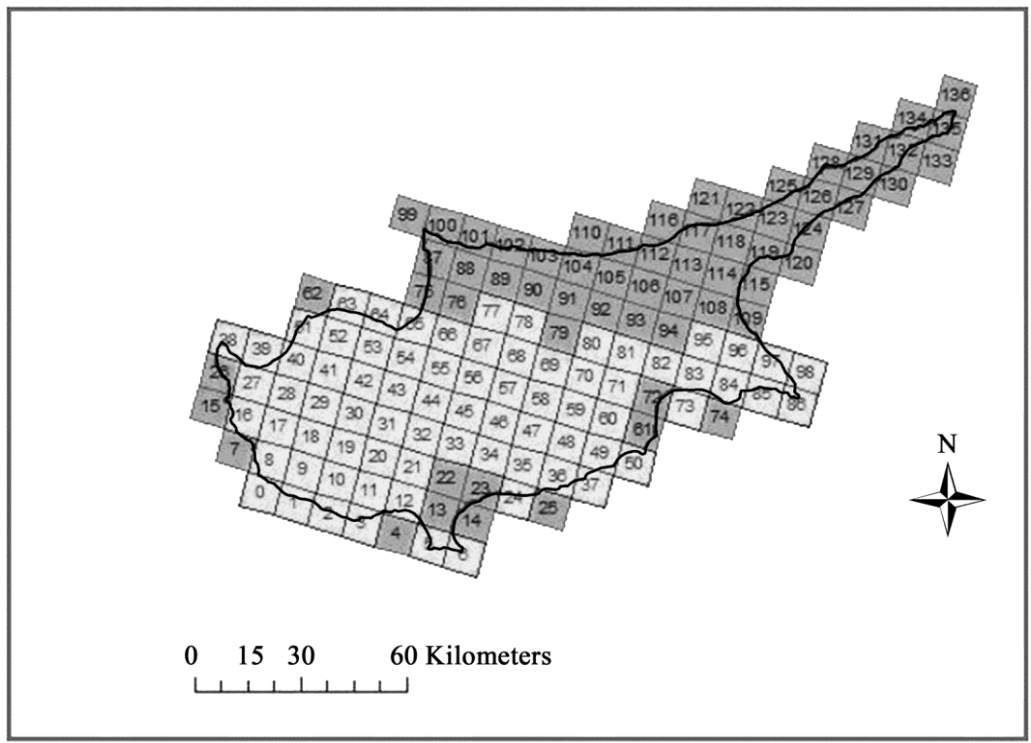

Figure. 1. Study area. In light grey, bird data grid cells used in the analysis (see text for details)

*Road-related properties*

We used the Annual Traffic Census 2012 data to calculate basic traffic noise levels for each road and motorway of Cyprus. We assumed that basic traffic noise levels were fixed along the length of each road. For the calculations the "Calculation of Road Traffic Noise" software (CRTN) 1988 (http://resource.npl.co.uk/acoustics/techguides/crtn/) was used. We used a reference distance of 10 km from the nearest roadside based on vehicle flow, speed, traffic composition, gradient and road surface. Calculations were done as follows:

- Time period: $L_{10}$ hourly dB(A)
- Total Vehicle Flow: Veh/Hour
- Speed: Use of maximum limit per road category (110km/h for motorways/highways, 80km/h for the rest of the roads)
- Gradient: 0% was assumed for every road since there were no data available
- Road surface: impervious

We overlaid the road map to the bird data grid cell, and for every cell, the noise level was calculated at the individual road segment and the average per cell was taken. For example, in the case of cell number 1 (Fig 1) which was intersected by 27 road segments, we calculated basic traffic noise level for all 27 segments and estimated its noise level as the average of noise for its constituent segments.

In addition, we calculated the following properties for every grid square: total road density and total road area extent, motorway and non-motorway length and roadless area i.e. areas with no roads at least 1 km away from the nearest road (BfN 2008). In order to characterise the anthropogenic penetration of landscapes from a geometric point of view, we calculated Landscape Division, Splitting Index and Effective Mesh Size metrics (Jaeger 2000) for every grid cell, using V-LATE 2.0 - beta extension for ArcGIS 10 (Lang and Tiede 2003). Table 1 shows all the properties and variables used.

Table 1. Hierarchical Variance Partitioning Results

| Variable | Description | Range | % of variance explained | After 100 randomizations Obs | Z.score |
|---|---|---|---|---|---|
| 1. Road length (Road_lengt) | Length of roads in km | 1 - 486 km | 7.8 | 1.26 | 0.62 |
| 2. Total road area/extent (Tot_Area) | Total area of roads in sq.km | 1- 100 km2 | 49.26 | 8.38 | 6.48* |
| 3. Motorway length (Motorway_l) | Length of motorways in km | 0 - 39 km | 1.69 | 0.27 | -0.48 |
| 4. Division | As the probability that two randomly chosen places in the landscape under investigation are not situated in the same undissected area | 0 - 94 | 11.10 | 1.88 | 0.89 |
| 5. Mesh | size of the areas when the region under investigation is divided into S areas (each of the same size At =S) with the same degree of landscape division | 1 - 100 | 9.00 | 1.5 | 0.73 |
| 6. Roadless area (RoadlessAr) | areas with no roads at least 1 km away from nearest road | 0 - 90 km2 | 9.94 | 1.65 | 0.99 |
| 7. Average Noise (AverNoise) | Average noise level in db for all constituent segments within a cell based on CRTN software | 0 - 71db | 3.60 | 0.59 | -0.36 |
| 8. Length of Roads with Noise (RoadwNoise) | Length of roads with calculated noise levels | 0 - 72 km | 7.57 | 1.24 | 0.24 |

*significant at 95%

*Mapping Noise and Fragmentation Hotspots*

Using the Jenks Natural Breaks Classification Method (Jenks 1967) we classified the 75 squares in the study area in terms of noise and fragmentation, as measured by landscape division into three classes: 'low', 'moderate' and 'high' impact (Table 1). We then mapped the combined impacts using a simple matrix approach (Fig 2) which resulted in five impact categories: very low, low, moderate, high and very high.

Fig. 2. Noise and Fragmentation Combined Impact Matrix

*Statistical Analysis*

We used Principal Component Analysis (PCA) to evaluate correlations between environmental variables independently of the species data. We sought to quantify the relative importance of road attributes, noise related parameters and fragmentation parameters (by-products of roads configuration) associated with species richness as deduced from Birdlife data within the mapping/recording scheme. We implemented Hierarchical Variance Partitioning (HVP) statistical modelling to account for the contribution of each explanatory variable to the total variance (Mac Nally 2002). HVP is a statistical framework that is capable of handling correlated independent variables, whilst providing a reliable ranking of predictor importance of each variable (Mac Nally 2002). Variance partitioning is calculated from the Akaike (AIC) weights of each explanatory variable and it is based upon the number of times that a variable was significant in all possible combinations of the explanatory variables. We conducted variance partitioning using the 'all.regs' and 'hier.part' functions in the hier.part package in R (R Development Core Team 2014). We performed 100 randomisation runs to extract confidence intervals for every variable.

**Results**

*Noise & Fragmentation*

Road traffic noise is higher within the coastal areas of the island of Cyprus which is home to many large towns with high motorway density (noise range from 60 to 71db). Road traffic noise on the mountainous areas was 'very low' to 'zero' (e.g. Troodos range) (Fig 2). Fragmentation is 'high' (value range 33 - 66) and 'very high' (value range 66 - 94) in most of the cells examined, even in mountain areas, which is a result of the dense road network on the island (Zomeni and Vogiatzakis 2014). Out of a total of 75 grid cells, two cells were of very low impact, 11 of low, 14 of moderate, 21 of high and 27 of very high combined impact (Fig 3). Cells with very high combined impact were in urban and peri-urban areas with many high impact cells in the south and centre of the island including the south-southwest parts of the Troodos massif. Overlaying the combined impact map with Natura 2000 Special Protection Areas (SPAs) demonstrate that there are 11 SPAs which are entirely or partly within areas with high and very high combined impact viz (1) Pentaschinos River (CY6000008), (2) Kosiis - Pallourokampou (CY 6000009), (3) Xyloyrikou Valley (CY 5000008), (4) Faros Kato Paphou (CY4000013), (5) Madari-Papoutsa (CY 2000015), (6) Agia Thekla - Liopetri (CY 3000009), (7) Troodos National Forest Park (CY 5000004), (8) Cha-Potami (CY5000010), (9) Paralimni Lake (CY3000008), (10) Gavo Gkreko (CY 3000005), (11) Atsa – Agios Theodoros (CY 2000014).

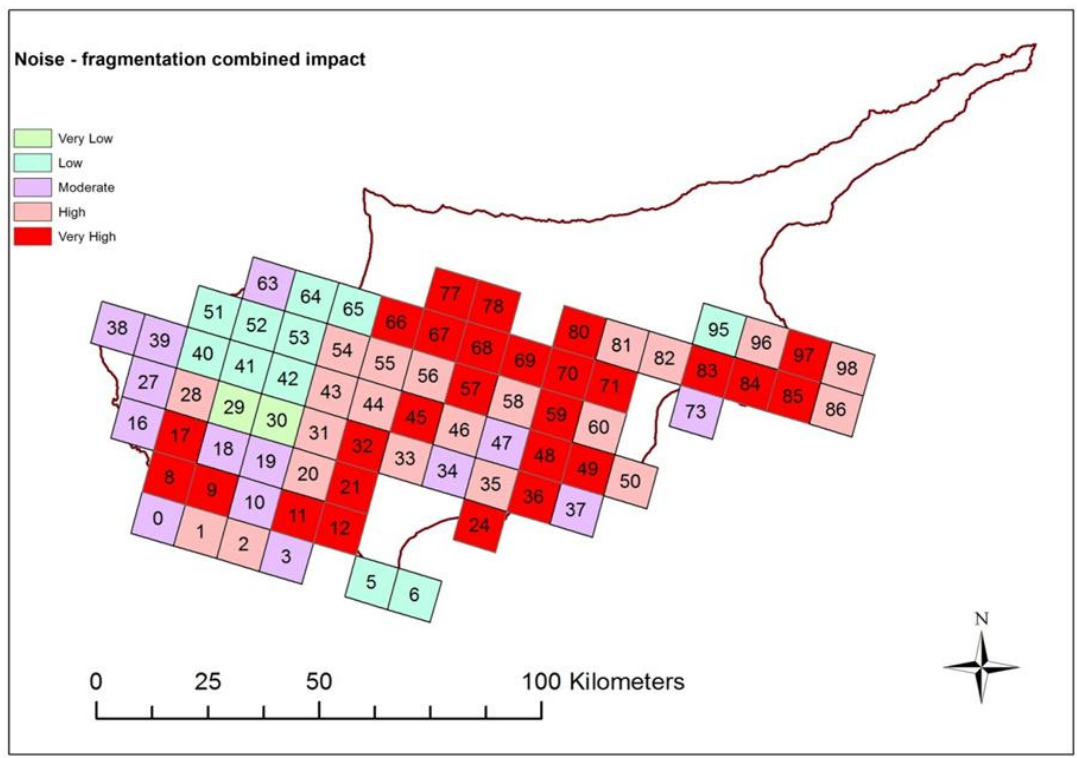

Fig. 3. Combined impact of noise and fragmentation in every grid cell evaluated

*Statistical Analysis – Hierarchical Partitioning*

The results of PCA (Fig 4) indicate that there was a very high correlation between split and length of roads with noise. We therefore retained only the latter for further HVP analysis. Results from HVP indicated that road-related properties (total road extent and road length) accounted for a combined 59% of variation, followed by fragmentation related properties (30% of variation) and noise properties (11% of variation). In terms of individual covariates, the predominant covariates of species richness were areas of road (Fig. 5) accounting for ca.50% of variation. This was followed by division (11.1%), mesh (9%) and roadless area (9.9%). The statistical significance of the area is confirmed also by z score following 100 randomisations.

Table 2. Impact Classes for Noise and Fragmentation

|  | Noise (dB) | Division |
|---|---|---|
| **Classes** | | |
| Low | 0-1l | 0 - 33 |
| Moderate | 10-65 | 33-66 |
| High | 65 -71 | 66 - 94 |

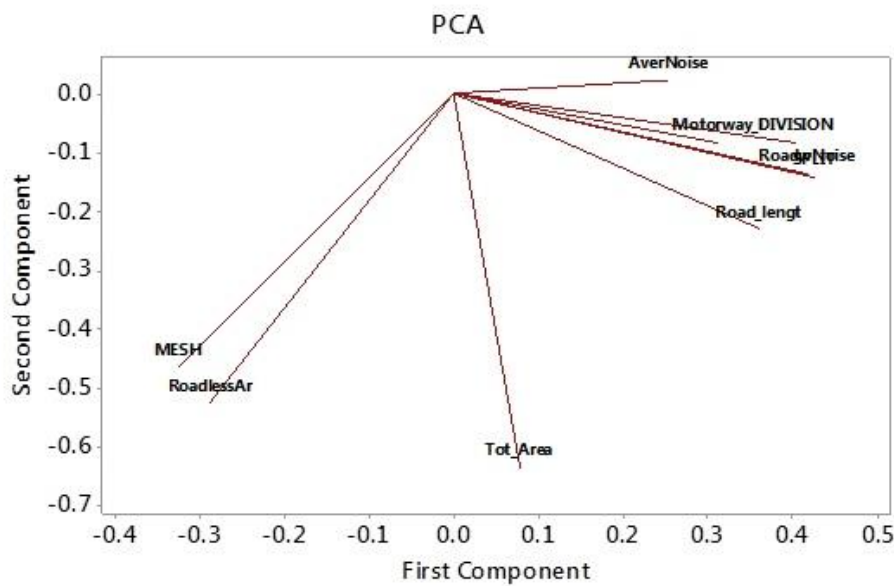

Fig. 4. PCA results of road related variables employed (for variables abbreviations see Table 1)

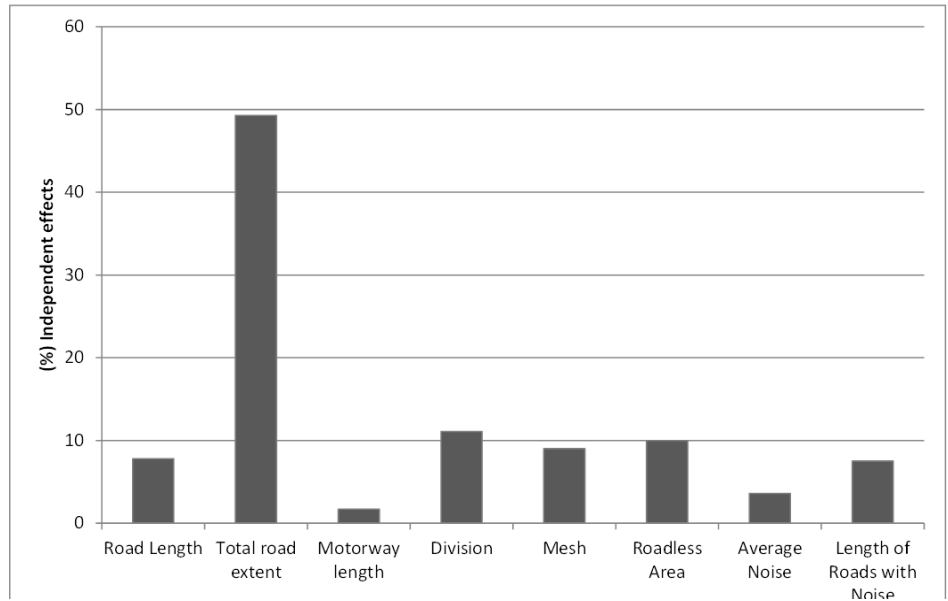

Fig. 5. The independent contribution (given as % of variance explained) of the variables employed for the total bird richness as estimated from hierarchical partitioning

**Discussion**

This study sought to evaluate the effect of a number of measurable road-related variables i.e. road network attributes and road-induced fragmentation and noise on bird species richness in Cyprus. This is the first time that a study of such a scale has taken place on the island. We based the methodology tested on widely employed tools and techniques for quantifying noise and fragmentation.

Noise levels are predictably higher in cells containing major urban centres and the ones adjacent to them (66-71dB). Although high noise levels may have negative effects on habitat quality, there are several factors that determine these effects, one of those being the inherent ambient noise in bird habitats (Ortega 2012). Nevertheless, evidence suggests that birds can adapt to noise through shifts in vocal amplitude, evolution of bird calling and by movement to other positions (Brumm and Slabbekoorn 2005; Ortega 2012). Based on typical noise levels in a quiet rural area, traffic noise levels approaching 50-60 dB(A) are assumed to begin to interfere with acoustic communication, with 60 db(A) considered a critical noise level (Dooling and Popper 2007). According to our findings 46 cells were well over that range with a maximum value of 71dB. In some areas e.g. industrial areas outside major urban centres, the values calculated in this study are probably an underestimate since there are significant additional noise sources of anthropogenic origin. In addition, in high upland areas such as the Troodos Massif, the lack of information on road gradient for model parameterisation might have also influenced our results. However, the values calculated using CRTN software are in line with the published noise maps for Cyprus which are limited to urban and peri-urban areas (Department of Environment 2013). Moreover, critical noise levels are seen as indicative given that their impacts are species specific. Helldin et al. (2013) developed a method to calculate the effective habitat loss due to traffic noise for two important bird areas in Sweden. They concluded that the impact from traffic noise represents an effective loss of 0.02-1.7% of the total area of the selected habitat types.

Fragmentation, measured with three of the most widely used metrics demonstrated that even mountainous areas are highly impacted, reflecting the existing dense road network. The combination of noise with fragmentation maps identified many areas with high and very high impact (hotspots), containing 11 SPAs under the Natura 2000 network of protected areas which present significant risk for Cyprus's avifauna, where mitigation actions should be taken. This corroborates with the results from other studies on the island on the pervasiveness of the road network on protected areas (Zomeni and Vogiatzakis 2014; Mammides et al. 2015).

Hierarchical partitioning results indicate that road properties are more important than road-induced fragmentation and noise in explaining variation in bird diversity, accounting for 59% of the total variation. Road area (extent) was more important than road length, which in our case is equivalent to road density, since the reference area was the recording grid's cell size (i.e. 10km x 10km). Usually road density is the most common property used in the evaluation of road related studies on biodiversity. Our results suggest however that a complete picture of the extent of roads (i.e. density and area) should be considered when we evaluate road impacts on biodiversity.

Fragmentation-related properties combined accounted for 30% of the variation in bird species richness with *degree of landscape division* being the most important among them followed by *roadless area* and *effective mesh size*. The effects of fragmentation on birds have been well documented (Stephens et al. 2004; Lampila et al. 2005; Barlow et al. 2006). At global scales, these effects seem to be mediated by latitude and may be specific to functional group (Bregman et al. 2014). There are many metrics proposed for evaluating road-induced fragmentation among which are the three indices employed in this study, which are widely used (Moser et al. 2007) and have been advocated as appropriate measures of the geometrical properties of an area following fragmentation (Jaeger 2000).

The importance of roadless areas for conservation is well-documented (Selva et al. 2011; Psaralexi et al 2017): they are areas of utmost importance for biodiversity conservation, ecosystems processes and connectivity, as well as maintaining ecological integrity (DeVelice and Martin 2001; Loucks et al. 2003; Crist et al. 2005). As a result, they usually contain more species, including those with large area requirements (carnivores), core area species and species sensitive to human disturbance (Haskell 2000; Watkins et al. 2003; Angelstam et al. 2004; Blake et al. 2008; Chen and Roberts 2008). In addition, they may serve as buffers to parasites, diseases and invasive species (Strittholt and DellaSala 2001; Allan et al. 2003; Gelbard and Harrison 2003; Holdsworth et al. 2007; Von der Lippe and Kowarik 2007). Zomeni and Vogiatzakis (2014) have identified 222 roadless areas scattered across the island covering only 4.5% of its surface. Road density is often used as a proxy for landscape impermeability for large carnivores (Carroll et al. 2003) and *road total area* is an aspect of road network extent that gives more accurate information to assess road proximity. Both proximity to roads and high road density influence reproductive success for a range of bird groups and guilds (Catchpole and Phillips 1992; Donazar et al. 1993; Reijnen and Foppen 1994) and affect the composition of neighbouring bird communities (Glennon and Porter 2005). The main response by birds in the vicinity of human infrastructure is either avoidance or a reduced population density with the effect of infrastructure on bird populations extended over distances up to about 1 km, depending on habitat structure (open vs dense) and guild (Benitez-Lopez et al. 2010). The adverse effects of road density on birds in Cyprus have been recently reported by Mammides et al. (2015) owing to both direct and indirect effects.

Results in this study were influenced by the (lack of) large available datasets (Moustakas and Evans 2017; Evans et al. 2014) and thus to that end the interpretation of the results derived here should account for this fact (Moustakas 2017). Bird species data were available at a coarse scale (10 km x 10km) but this is currently the only dataset available for the whole island, which despite some gaps, follows a consistent monitoring scheme used widely in Europe. The resolution of species data also dictated the aggregation of the variables used in statistical analysis at the same resolution. Landscape metrics in particular are known to be scale and guild specific. Although there are many ways to model road traffic noise, the CRTN software has been widely employed in noise related studies (Li et al. 2002; O'Malley et al. 2009) and includes the most important parameters accounting for noise. It is openly distributed and easy to use without recourse to complex modelling. Conclusions in bird studies are generally confounded by the inclusion of habitat-related variables and/or landscape heterogeneity along with road properties (Smith et al. 2011). In this case, given the importance of habitat and landscape attributes on bird diversity, we sought to investigate the role of road related properties independently of intrinsic habitat and landscape properties, and explain variation rather than predict species richness.

**Conclusion**

There is an increased interest in the effects of roads on landscape structure, habitat and species loss particularly in protected areas (Ament et al., 2008, Garriga et al., 2012) and therefore it is important to identify and buffer any negative effects of roads. However, there is limited knowledge on whether the decrease in birds' resilience is affected most by direct or indirect consequences. If this could be determined then it would be possible to manage road networks to the benefit of bird populations to create habitats (Forman 2000), increase availability for food resources (Lambertucci et al. 2009) and even connect habitat patches (Huijser and Clevenger 2006; Reijnen and Foppen 2006). The purpose of this research was to reveal any noteworthy impacts of roads, including extensive fragmentation and traffic noise, on bird populations in Cyprus, and contribute towards improved transport infrastructure planning and enhance public awareness on the subject.

**Acknowledgements**

We would like to thank Birdlife Cyprus for providing the breeding bird dataset and the Department of Public Works at the Ministry of Transport, Communication and Works of Republic of Cyprus for the Annual Traffic Data Census provision.